\documentclass[aps,prl,superscriptaddress,reprint]{revtex4-1}
\usepackage{blindtext}
\usepackage{amsfonts,amssymb}
\usepackage[utf8]{inputenc}
\usepackage{graphicx}
\usepackage{color}
\usepackage{hyperref}
\hypersetup{colorlinks,allcolors=blue}
\usepackage{filecontents}

\font\cmss=cmss10
\font\cmsss=cmss10 at 7pt
\font\manual=manfnt

\newcommand{\bi}{\begin{itemize}}
\newcommand{\ei}{\end{itemize}}

\newcommand{\bea}{\begin{eqnarray}}
\newcommand{\eea}{\end{eqnarray}}
\newcommand{\be}{\begin{equation}}
\newcommand{\ee}{\end{equation}}
\newcommand{\ben}{\begin{eqnarray*}}
\newcommand{\een}{\end{eqnarray*}}
\newcommand{\bem}{\begin{pmatrix}}
\newcommand{\eem}{\end{pmatrix}}
\newcommand{\bl}{\begin{align}}
\newcommand{\el}{\end{align}}
\newcommand{\beg}{\begin{gather}}
\newcommand{\eeg}{\end{gather}}

\newcommand{\cO}{\mathcal{O}}

\newcommand{\IH}{\mathbb{H}}

\renewcommand{\a}{\alpha}
\renewcommand{\b}{\beta}
\renewcommand{\d}{\delta}
\newcommand{\e}{\epsilon}
               
\newcommand{\g}{\gamma}

\renewcommand{\l}{\lambda}
\newcommand{\m}{\mu}
\newcommand{\n}{\nu}

\renewcommand{\r}{\rho}

\newcommand{\D}{\Delta}

\renewcommand{\L}{\Lambda}
\renewcommand{\O}{\Omega}

\newcommand{\half}{\frac{1}{2}}

\def\dbend{\lower3.5pt\hbox{\manual\char127}}

\def\IL{\relax{\rm I\kern-.18em L}}
\def\IH{\relax{\rm I\kern-.18em H}}
\def\rlx{\relax\leavevmode}

\def\ZZ{\rlx\leavevmode\ifmmode\mathchoice{\hbox{\cmss Z\kern-.4em Z}}
 {\hbox{\cmss Z\kern-.4em Z}}{\lower.9pt\hbox{\cmsss Z\kern-.36em Z}}
 {\lower1.2pt\hbox{\cmsss Z\kern-.36em Z}}\else{\cmss Z\kern-.4em
 Z}\fi}

\def\TrH#1{ {\raise -.5em
                      \hbox{$\buildrel {\textstyle  {\rm Tr } }\over
{\scriptscriptstyle \CH _ {#1}}$}~}}

\begin{document}

\title{Quantum Weyl Invariance and Cosmology}
\author{Atish Dabholkar}

\affiliation{
{\it International Centre for Theoretical Physics, ICTP-UNESCO, Strada Costiera 11, Trieste 34151 Italy }}
\affiliation{{\it Sorbonne Universit\'es, UPMC Univ Paris 06, CNRS UMR 7589, LPTHE, F-75005, Paris, France}
}

\date{November 2015}

\begin{abstract}
Equations for cosmological evolution are formulated in a Weyl invariant formalism to take into account possible Weyl anomalies. Near two dimensions, the renormalized cosmological term leads to a nonlocal energy-momentum tensor and a slowly decaying vacuum energy. A natural generalization to four dimensions implies a quantum modification of Einstein field equations at long distances. It offers a new perspective on time-dependence of couplings and naturalness with potentially far-reaching consequences for the cosmological constant problem, inflation, and dark energy. 
\end{abstract}

\maketitle

To define a path integral  over metrics in a quantum theory of gravity, one  must introduce a regulator. Since the metric itself is a dynamical field, it is not clear in which metric to regularize and renormalize the theory, and how to ensure that the resulting answer is coordinate invariant and background independent. For this purpose it is  convenient to  enlarge the gauge symmetry to include Weyl invariance in addition to general coordinate invariance. This can be achieved by introducing  a Weyl compensator field and a fiducial metric which scale appropriately  keeping the physical metric Weyl invariant. The number of degrees of freedom remains the same  upon imposing Weyl invariance. 
 The path integral can now be regularized and renormalized using the fiducial metric. 
 
A Weyl-invariant formulation has an important conceptual advantage because it separates scale transformations from coordinate transformations. The path integral can be regularized maintaining coordinate invariance  at the quantum level. Weyl invariance can have potential anomalies in the renormalized theory but since it is a  gauge symmetry  all  such anomalies must cancel. Coordinate invariance of the original theory then becomes  equivalent to  coordinate invariance plus  \textit{quantum} Weyl invariance of the modified theory. This procedure  is well-studied in two dimensions where the Liouville field plays the role of the Weyl compensator and quantum Weyl invariance implies nontrivial scaling exponents. 

There are both theoretical and phenomenological motivations to develop  a Weyl-invariant formulation  of gravity in higher dimensions, especially in the context of cosmology.
Our chief theoretical motivation  is to formulate the cosmological constant problem \cite{Weinberg:2000yb} in a manifestly gauge invariant way. The problem  is usually stated in the language of effective field theories as a `naturalness problem' analogous to the Higgs mass problem in electroweak theory or the strong-CP problem in quantum chromodynamics. The cosmological constant is the coupling constant of the identity operator added to the effective action. Since the identity has dimension zero,  the cosmological constant term is the most  relevant operator and  should scale  as $M_{0}^{d}$ in $d$ space-time dimension where the ultraviolet cutoff scale $M_{0}$ is  at least  of the order of a TeV.  To reproduce the observed scale of the cosmological constant of the order of  an meV,  it is necessary to fine tune the  bare vacuum energy. 

This formulation of the cosmological constant problem is not entirely satisfactory. While the \textit{generation} of the cosmological constant  in the effective action depends only on short-distance physics, its \textit{measurement} relies essentially on long-distance physics spanning almost the entire history of the universe. 
The physics of the cosmological constant thus spans  more than a hundred logarithmic length scales. Moreover, all scales are evolving in a cosmological setting,  and there is no preferred time for setting the cutoff  in a manner that respects  coordinate invariance. Thus, even to pose the cosmological constant problem properly,  it is desirable to develop a formalism that accesses  all time-scales  in a gauge-invariant fashion.

A chief phenomenological motivation is to explore the possibility of effective time variation of  vacuum energy.
There is a substantial body of cosmological evidence for a slowly varying vacuum energy   which  is believed to have been responsible for an inflationary phase of exponential expansion in the very early universe.   Observations of cosmic microwave background radiation indicate that the power spectrum  generated during inflation is not strictly scale-free but has  a slight  red tilt. This implies that vacuum energy was not strictly constant  but was slowly decaying during the inflationary era. Cosmological data also indicates that $69\%$ of present energy density is in the form resembling vacuum energy.  Time variation of  dark energy is not established observationally at present but  could be observed in  planned observations. Any theoretical insight into the  magnitude, equation of state, and  time dependence of dark energy is clearly desirable.

Slowly varying vacuum energy can be represented by a cosmological constant $\L$ to first approximation.  However, any time variation cannot be reintroduced simply by  making $\L$  time-dependent because that  would not be coordinate-invariant. A simple way to obtain time-dependent vacuum energy is to  represent it by a slowly-rolling condensate of a scalar field. This idea is central to most current models of varying vacuum energy. Such a slowly-rolling field is called the  `inflaton'  during the inflationary era and `quintessence'  during the present era. Models with scalar fields have the virtue of simplicity, but among the plethora of models none is particularly more compelling than the  others; and our understanding of many  questions of principle such as the  initial value problem or the measure problem is less than satisfactory.

The puzzles regarding the cosmological constant, inflation, and dark energy all concern the nature of slowly varying vacuum energy. Occam's razor suggests that perhaps   the essential underlying physics  is governed  by the same fundamental equations.   With these motivations, I develop  a Weyl-invariant formulation of quantum cosmology  to  explore  the possibility of  slowly evolving vacuum energy that does not rely on fundamental scalars. 

I start with a Weyl-invariant reformulation of classical  general relativity in $d$ spacetime dimensions by introducing a Weyl compensator field $\O$ and a fiducial metric $h_{\mu\nu}$.  Given a  UV cutoff $M_{0}$, the reduced Planck scale $M_{p}$ and the cosmological constant $\L$ 
correspond to  dimensionless `coupling constants' $\kappa^{2}$ and $\lambda$ defined by:
 \begin{equation}\label{dimless}
M_{p}^{d-2} := \frac{M_{0}^{d-2}}{\kappa^{2}}  \,  \qquad \Lambda :=   \lambda \kappa^{2} M_{0}^{2} \, .
\end{equation} 
 The gravitational action $I_{K}[h, \O]$ is given by
 \bea\label{WIaction}
\frac{M_{0}^{d-2}}{2\kappa^{2}}\int dx \, 
e^{(d-2)\Omega} [  R_{h} + (d-2)(d-1) |\nabla \Omega)|^{2} ] \, 
\eea
where all contractions  are using the metric $h$ and $dx := d^{d}x \,  \sqrt{-h}$.
The cosmological term is given by
\be\label{cosmo}
I_{\L}[h, \O]= - {M_{p}^{d-2} \L}\int dx \,  e^{d\O}= -\l M_{0}^{d} \int dx \,  e^{d\O}\, .
\ee

All terms are  coordinate invariant. 
Both $I_{K}$ and $I_{\L}$ are separately  invariant under Weyl transformations:
\be\label{Weyl}
h_{\m\n} \rightarrow e^{2\xi}h_{\m\n} \, , \qquad \O \rightarrow \O -\xi \, .
\ee
Consequently both $I_{K}$ and $I_{\L}$ satisfy the  Ward identities for coordinate invariance:
\be\label{Diff-WT}
\nabla^{\n} ( \,\frac{-2 \, \d I_{a}}{\sqrt{-h} \, \d h^{\m\n}})  -    \frac{1}{\sqrt{-h}}\frac{\d I_{a}}{\d \O}\, \nabla_{\m} \O   \,  \equiv 0 \quad (a= K, \L)\,  ,
\ee
and for Weyl invariance:
\be\label{Weyl-WT}
 h ^{\m\n}  ( \,\frac{-2 \, \d I_{a}}{\sqrt{-h} \, \d h^{\m\n}})-    \frac{1}{\sqrt{-h}}\frac{\d I_{a}}{\d \O}    \,  \equiv 0 \quad (a= K, \L)\, .
\ee
The physical metric $g_{\m\n} := e^{ 2\O} h_{\m\n}
$ is Weyl invariant. In the `physical' gauge we have $\O=0$ and $h_{\m\n} =g_{\m\n}$ and (\ref{WIaction}) reduces to the Einstein-Hilbert action.  

Consider a  homogeneous and  isotropic universe described by a spatially flat Robertson-Walker  metric with scale factor $a(t)$,  filled with a perfect fluid of  energy density $\rho$ and pressure $p$.  The classical evolution of the universe is governed by the first Friedmann equation
\bea\label{Friedmann1}
H^{2} =  \frac{ 2 \kappa^{2} \r } {(d-2)(d-1) M_{0}^{d-2}} \, 
\eea
and the conservation equation
\be\label{Conservation}
\dot{\rho}= - (d-1)  (p + \r) H \, .
\ee
For a  perfect fluid with a barotropic equation of state  $p = w\r$, the solutions to (\ref{Friedmann1}) and (\ref{Conservation}) are given by 
\be\label{density}
\rho(t)  = \rho_{*} (\frac{a}{a_{*}})^{ -\g} \, , \qquad  a(t) = a_{*}(1 + \frac{\gamma}{2} H_{*} t) ^{\frac{2}{\g}} \, ,
\ee
where $\r_{*}$, $H_{*}$, $a_{*}$ are  the initial values of various quantities at $t=0$, and  $\g := (d-1) (1 + w)$.
For the classical  tensor of the cosmological term,  $\r_{*} = \l_{*}M_{0}^{d}$, $w=-1$, and $\g =0$. As  $\g \rightarrow  0$, the solution approaches nearly de Sitter spacetime with nearly exponential expansion and nearly constant density. Note that the cosmological evolution equations depend analytically on $d$, so one can `analytically continue' the FLRW cosmologies.

Weyl invariance has potential  anomalies at the quantum level.   
To gain intuition about these anomalies, we first consider  spacetime near two dimensions, $d = 2 +\e$. To order $\e$, the total action $I$ without matter  is given by
\bea\label{gravityaction2d}
 \frac{q^{2}}{4\pi}\int dx \big( \frac{R_{h}}{\e}\,+\,| \nabla \Omega|^2 + R_h \Omega   -\frac{{4\pi\l}M_{0}^{2}} {q^{2}} e^{2\Omega}\big) 
\eea
where the coupling constant $q$ defined by
\be 
{q^2} :=\frac{2\pi \epsilon}{\kappa^2}\, 
\ee
is held fixed as $\e \rightarrow 0$. 
With $\chi :=q\, \O $ and $\mu = \l M_{0}^{2}$, and  ignoring the first term which depends only the fiducial metric, (\ref{gravityaction2d}) is precisely  the two-dimensional Liouville action with  background charge $q$:
\be\label{timelike}
{I[{\chi}]= \frac{1}{4\pi }\int dx \,\left( | \nabla \chi|^2 + q \, R_h \,\chi  -   \, 4\pi \mu\,  e^{{2 \beta \chi} }  \right)} \, .
\ee
The field $\chi$ is sometimes called the `timelike' Liouville field because the kinetic term has a wrong sign, as expected for the conformal factor of the metric. Classical Weyl invariance implies $\b = 1/q$,  but this relation receives quantum corrections because the  operator $e^{{2 \beta \chi} }$ is a composite operator with short-distance singularities. 
%
It can be renormalized treating $\chi$ as a free field \cite{Polyakov:1981rd}  with the Green function $G_{2}$ of the Laplacian $\D_{2}$:
\be\label{Green2}
 \Delta_{2}^{x}\, G_{2}(x, y) = \delta_{2}(x, y) \, .
\ee 
There is  a short-distance divergence arising from self-contractions  which combine into an exponential of the  coincident  Green function $G_{2}(x, x)$. This divergence can be regularized by using a heat kernel with a short-time cutoff. Renormalization then consists in subtracting a logarithmically divergent term from the regularized $G_{2}(x, x)$. This procedure is manifestly local and coordinate invariant.  In two dimensions, any metric is conformal to the  flat metric $\eta_{\m\n}$:
$h_{\m\n}=e^{2\Sigma} \eta_{\m\n}$. The renormalized  operator 
$
\cO_{h}(x) := [e^{{2 \beta \chi} }]_{h}
$
depends on the fiducial metric used for regularization and 
satisfies
\be\label{heta}
 \cO_{h}(x) = e^{-2 \b^2   \Sigma(x)}\,\cO_{\eta}(x)\,  .
\ee 
The scalar $\Sigma$ is a  \textit{nonlocal} functional  of the metric:
\be\label{sigma2}
{\Sigma}[h] (x) :=  \frac{1}{2}\int dy \, G_{2}(x, y) R_{h} (y) \, .
\ee
The defining  property of $\Sigma$ is its  Weyl transformation
\be\label{sigmatransform}
\Sigma [h] \rightarrow \Sigma [h] + \xi  \quad \textrm{when} \quad h_{\m\n} \rightarrow e^{2\xi} \,  h_{\m\n} \, .
\ee
Hence the renormalized operator  has anomalous dimension $2\beta^{2}$.  The total Weyl variation is  given by
\be
\cO_{h} \rightarrow e^{-2{(\b q + \b^{2}) \xi}} \cO_{h} \, .
\ee

The renormalized cosmological term is
\be\label{rencosmo}
I_\L= -\m\,\int dx \, \cO_{h}(x)  \, .
\ee
Weyl invariance  of $I_{\L}$ now implies
\be\label{anomalous}
2\b q + 2\b^{2} = 2 \quad \textrm{or} \quad q = \frac{1}{\beta} -\b \,  \,\, \textrm{and}  \,\, \b = \frac{1}{q} +  \ldots.
\ee

It is illuminating to interpret these results in terms of a quantum effective action. In operator formalism, the renormalization of the cosmological operator corresponds to normal ordering in the conformal vacuum defined using  the  Klein-Gordon inner product in the metric $\eta_{\m\n}$.  Evaluating  the operator equation (\ref{heta}) around a classical background field $\Omega$ in the conformal vacuum  and by using   (\ref{anomalous}) and (\ref{rencosmo}), we obtain the quantum effective action that replaces the classical action (\ref{cosmo}):
\be\label{qaction}
I_{\L}= -\m \int dx\, e^{2\Omega}\, e^{-2\b^{2}(\O + \Sigma)} \, .
\ee
The term depending on $\Sigma$ encapsulates the anomalous dimension of the operator and is nonlocal. The corresponding momentum tensor  is 
\bea\label{quantum stress tensor}
{T}^{\L}_{\m\n}(x)=-\m \, (1 -\beta^{2})   h_{\m\n} \cO_h(x) + 2\m\b^2  S_{\m\n}(x)
\eea
where $S_{\m\n}$ is  nonlocal and traceless:
\bea\label{Stensor}
&&S_{\m\n}(x)= \int dy \, \Big[ \nabla^x_\m \nabla^x_\n- \half h_{\m\n} \, \nabla^x \cdot \nabla^x \Big] G_{2}(x, y ) \, \cO_h(y )  \nonumber\\
&&+  \int dy\,  dz
\Big[\nabla^x_{(\m} G_{2}(x, y )\, \nabla^x_{\n)} \, G_{2}(x, z) \\
-&&\half h_{\m\n}(x)\,h^{\a\b}(x)\, \nabla^x_{\a}\, G_{2}(x, y )\, \nabla^x_{\b} \, G_{2}(x, z) \Big] \, \cO_h(y )\,R_h(z) \, . \nonumber
\eea
The nonlocality of the momentum tensor  reflects the nonlocality of the   action (\ref{qaction}) and is in line with the interpretation of anomalies as the effect of regularization that  cannot be removed by local counterterms. Since $(\Omega + \Sigma)$ is a Weyl-invariant scalar, the action (\ref{qaction}) satisfies both Ward identities (\ref{Diff-WT}) and (\ref{Weyl-WT}) and the quantum momentum tensor (\ref{quantum stress tensor}) is conserved. 

The total quantum  action in physical gauge is given by
\be\label{qEH}
I[g] = \frac{M_{p}^{2}}{2}\int d^{2+\e}x \sqrt{-g}\left[  R_{g} - 2\L e^{-2\b^{2} \Sigma [g]}\right] \, .
\ee
The right-hand side of Einstein field equations now has a nonlocal momentum tensor. Correspondingly the Friedmann equations are replaced by  new \textit{nonlocal integro-differential equations} for cosmological evolution. One might worry that the nonlocality would lead to ghosts and causality violations. However, one must use the in-in effective action in the Schwinger-Keldysh formalism and not the in-out effective action. The corresponding boundary conditions naturally lead to retarded Green functions instead of Feynman propagators, thus ensuring  causality of the quantum cosmological evolution. 

For a spatially-flat Robertson-Walker metric,  the second term in $S_{\m\n}$ vanishes and only time derivatives contribute to the first term.  Consequently, the quantum momentum tensor evaluated on the Robertson-Walker ansatz turns out to be local, corresponding to  a barotropic fluid with 
$w_{\L} = -1+ 2\beta^{2}$ or $\g = 2\beta^{2}$.  Using (\ref{density}) we arrive at a dramatic conclusion that the vacuum energy decays  and the effective coupling constant $\l$ evolves as the universe expands: 
\be\label{decay}
\rho(t) =  \r_{*} (\frac{a}{a_{*}})^{-2\beta^{2}} \quad \textrm{or} \quad \l(t) =  \l_{*} (\frac{a}{a_{*}})^{-2\beta^{2}}  \,  .
\ee
This provides a dynamical mechanism for the relaxation of the `cosmological constant'. 

One expects  that a  composite operator like the determinant of the metric will have anomalous dimension even in four dimensions. The quantum action (\ref{qEH}) suggests a natural generalization  to four dimensions. Consider the  Weyl-covariant quartic operator  \cite{Fradkin:1982xc, Riegert:1984kt, Paneitz:2008} defined by
\be\label{riegert}
\Delta_{4}:= \nabla^{4} + 2 \, R^{\m\n} \, \nabla_{\m}\nabla_{\n} +\frac{1}{3} (\nabla_{\m} R)\, \nabla^{\m} - \frac{2}{3} R\, \nabla^{2}
\ee
and the corresponding Green  function $G_{4}$  satisfying:
\be\label{Green4}
 \Delta_{4}^{x}\, G_{4}(x, y) = \delta_{4}(x, y) \, .
\ee 
Then $\Sigma$ which transforms as in (\ref{sigmatransform}) is given by
\bea
&{\Sigma}[h] := \frac{1}{4}\int dy \, G_{4}(x, y) \, F_{4} (y) \, ; \label{sigma4}\\
F_{4} &:= R_{\m\n\a\b} R^{\m\n\a\b} - 4 R_{\m\n}R^{\m\n} + R^{2}-\frac{2}{3} \nabla^{2} R \, . \label{F4}
\eea
Using these ingredients I propose  a generalization of the Einstein-Hilbert action in the physical gauge: 
\be\label{newaction4}
I[g] = \frac{M_{p}^{2}}{2} \int d^{4}x \sqrt{-g}\left[ e^{-\g_{K} \Sigma[g]} R_{g} - 2\L\, e^{-\g_{\L} \Sigma [g]}\right]
\ee
where $\g_{K}$ and $\g_{\L}$ are the  anomalous dressings. 

The resulting equations of motion  are nonlocal and the momentum tensor has a complicated expression. Remarkably,  for the Robertson-Walker metric, the momentum tensor simplifies again describing  a barotropic perfect fluid with $w = -1 + \gamma/3$ and $\gamma = \gamma_{\L} -\gamma_{K}$. 
Thus vacuum energy decays slowly for small positive $\g$. The slow-roll parameters are of order $\g$ \cite{Dabholkar:2015}.

The ansatz (\ref{newaction4}) is motivated by the considerations of renormalization and quantum Weyl invariance. It would be important to compute the anomalous gravitational dressings in four dimensions from first principles.  Even in two dimensions,  such a  computation is nontrivial. The use of free Green function (\ref{Green2}) is justified ultimately by exact results obtained using conformal bootstrap. Analogous methods are not available in four dimensions, but we expect that semiclassical  computations should be feasible and reliable on cosmological scales.  From general renormalization group considerations, we expect non-vanishing anomalous  dressings for any composite operator and the corresponding metric dependence through $\Sigma$.  Since anomalous dressings in general can be scale dependent, they could be  arbitrary functions of $\Sigma$ \cite{Dabholkar:2015}. 

In Liouville theory, an operator  $\cO_{i}$ of mass dimension $\D_{i}$ can be coupled  in a Weyl invariant way with an action
\be
I_{i} = -\l_{i} M_{0}^{2-\D_{i}}\int dx \, \cO_{i} \, e^{(2-\D_{i})\O}e^{-\g_{i}(\O + \Sigma)} \, .
\ee
Here $\gamma_{i}$ is the  `anomalous gravitational dressing' which is the anomalous dimension of the composite operator $[\cO_{i} \, e^{(2-\D_{i} -\g_{i})\O}]$. For the identity operator, $\D_{\L} =0$ and $\g_{\L} =2\b^{2}$. Another example is the mass term for a fermion 
corresponding to the operator $\bar \psi\psi$ with $\D_{F} =1$:
\be\label{Dirac}
I_{F} = - \l_{m} M_{0} \int dx  \left[ \bar \psi \psi e^{2\a q\O}\right] \, .
\ee
Weyl invariance implies
\be
2\a q + 2\a^{2} = 1 \, .
\ee
and the anomalous  dressing is $\g_{F}= 2\a^{2}$.

In four dimensions, one expects similar  anomalous dressings for matter. 
This has a striking consequence in an expanding universe: dimensionless ratios determined by naive mass dimensions can change with time. For example, for the 2d fermion mass we have
\be
\l_{m}(t) = \l_{m*}  (\frac{a}{a_{*}})^{-2\a^{2}} \, .
\ee 
This  anomalous time evolution is similar to the slow time decay of vaccum energy (\ref{decay}).  Even though the fermion mass and vacuum energy are of order one at the  cutoff scale $M_{0}$ in the beginning, their effective values today can be smaller  because of the anomalous time evolution. 

I propose a `{Cosmological Naturalness Principle}': ``\textit{If there is a very small dimensionless parameter occurring in nature, then the anomalous gravitational dressing of the associated operator in the effective action is such that  the smallness of the parameter is  a  consequence of its anomalous  time evolution in an expanding universe}.''   
Lack of evidence for supersymmetry thus far  could be an indication that the Higgs mass problem has a primarily cosmological explanation. This would allow for a higher scale of  supersymmetry breaking. It would be  interesting to compute the relevant gravitational dressings in the microscopic theory. Since gravitational dressings depend on the microphysics,  their cosmological manifestations would provide a useful IR window into the UV physics.

Even a tiny positive $\g$ would solve the cosmological constant problem with vanishingly small vacuum energy at late times.
More generally, the mechanism of vacuum energy decay could have far-reaching consequences for our understanding of inflation and dark energy.  Slowly decaying vacuum energy in the early universe with small $\gamma$ can drive slow-roll inflation without an inflaton field.  By itself,  it would lead to an empty universe, but with matter fields, one can imagine scenarios for a `graceful exit' into a hot big bang.
For example,  there can be a phase transition with large latent heat triggered by the mass-squared of a scalar field turning negative, to start a radiation-dominated hot big bang. 

Such `inflation without the inflaton'  driven by the dynamics of the Omega field could be called `Omflation'. 
It seems though that unless  the  vacuum energy is much smaller than the radiation energy \textit{after} the latest phase transition, it would dominate radiation too soon to be compatible with big bang nucleosynthesis. At late times $\gamma$ may be too small, of order $G\L$.  Since the omflaton is a gauge degree of freedom,  primordial scalar curvature perturbations would have to be generated by curvaton-like  scalars \cite{Lyth:2001nq}. 
 It thus remains to be seen whether one can construct a consistent evolution of the universe that incorporates omflation   followed by a big bang to terminate with a small remnant dark energy, and whether $\g$ can be large enough during this evolution in the microscopic theory.  At any rate,  with our novel mechanism for vacuum energy decay, it becomes a dynamical question.

The new action (\ref{newaction4}) implies a quantum modification of Einstein field equations at long distances and possibly observable deviations from the predictions of Einstein Gravity.  It also predicts mild time-dependence for dark energy today which could be detectable if the anomalous dressings are sufficiently large.   
Gravitational dressings of other fields can lead to slow variations of coupling constants over time which could be  constrained by observation and may possibly be detectable. A detailed analysis of the theoretical and observational implications will be presented in forthcoming publications \cite{Dabholkar:2015}.

\bibliography{/Users/atish/Documents/notes/Weyl/2dcosmo/weyl}
\bibliographystyle{apsrev}
\end{document}